\documentclass[conference]{IEEEtran}
\usepackage{graphicx}
\usepackage{url}
 \usepackage[numbers]{natbib}

\AtBeginDocument{%
  }

\begin{document}

\title{Governance Matters: Lessons from Restructuring the data.table OSS Project}

\author{
  \IEEEauthorblockN{Pedro Oliveira\IEEEauthorrefmark{1},
                    Doris Amoakohene\IEEEauthorrefmark{2},
                    Toby Hocking\IEEEauthorrefmark{2},\\
                    Marco Gerosa\IEEEauthorrefmark{1},
                    Igor Steinmacher\IEEEauthorrefmark{1}}
    \IEEEauthorblockA{\IEEEauthorrefmark{1}Northern Arizona University, Flagstaff, AZ, USA\\
                    \{pro32, marco.gerosa, igor.steinmacher\}@nau.edu}
    \IEEEauthorblockA{\IEEEauthorrefmark{2}\texttt{data.table} Project\\
                    Toby.Dylan.Hocking@usherbrooke.ca daa464@nau.edu}
}

\maketitle

\begin{abstract}
Open source software (OSS) forms the backbone of industrial data workflows and enterprise systems. However, many OSS projects face operational risks due to informal or centralized governance. This paper presents a practical case study of \texttt{data.table}, a high-performance R package widely adopted in production analytics pipelines, which underwent a community-led governance reform to address scalability and sustainability concerns. Before the reform, \texttt{data.table} faced a growing backlog of unresolved issues and open pull requests, unclear contributor pathways, and bottlenecks caused by reliance on a single core maintainer. In response, the community initiated a redesign of its governance structure. In this paper, we evaluated the impact of this transition through a mixed-methods approach, combining a contributor survey (n=17) with mining project repository data. Our results show that following the reform, the project experienced a 200\% increase in new contributor recruitment, a drop in pull request resolution time from over 700 days to under a week, and a 3x increase in contributor retention. Community sentiment improved around transparency, onboarding, and project momentum, though concerns around fairness and conflict resolution remain. This case study provides practical guidance for maintainers, companies, and foundations seeking to enhance OSS governance. 
%We argue that well-designed governance is an enabler of OSS reliability and developer engagement, supporting the long-term resilience of software supply chains.
\end{abstract}


\section{Introduction}

Open source software (OSS) has revolutionized the technology industry by enabling collaborative development and free distribution of code that powers everything from smartphone apps to mission-critical infrastructure systems, fostering innovation while reducing costs and vendor lock-in. Research shows that over 90\% of commercial software includes OSS components~\cite{nagle2024value}, and virtually all cloud platforms, programming languages, and data infrastructures build upon OSS foundations~\cite{conti2024creators}. OSS projects, such as Kubernetes, TensorFlow, and Linux, are not merely technical assets---they are central to enterprise IT, supply chains, and innovation strategies~\cite{eghbal2016roads}.

As OSS adoption accelerates, governance fundamentally contributes to project health and sustainability. Governance structures---how decisions are made, roles, and responsibilities distributed---directly influence projects' capacity to scale, onboard contributors, and remain responsive to user needs~\cite{goggins2021making, trinkenreich2023belong}. Research shows that mature governance models are associated with stronger contributor retention, improved diversity, and more resilient communities~\cite{steinmacher2018barriers}. Conversely, projects lacking formal governance often encounter bottlenecks, contributor turnover, and burnout among core maintainers~\cite{guizani2021long, linaaker2024sustaining, raman2020stress}.

Despite these known challenges, governance in OSS projects remains underexplored in academic literature. Much of the available guidance is grounded in gray literature sources, such as community blog posts~\cite{eghbal2016roads, neary2020governance}, foundation governance templates (e.g., templates from the Apache Software Foundation~\cite{apacheGovernance}, the Linux Foundation~\cite{todoGovernance}, the Eclipse Foundation~\cite{eclipseGovernance}, and guidance from the Open Source Initiative (OSI)~\cite{osiGovernance}), and practitioner-oriented handbooks~\cite{ospo2023ggi}. While these sources offer practical insight, they are rarely evaluated through systematic empirical methods. Therefore, the practical implications of governance structures remain under-assessed. This gap is troublesome, as governance breakdowns can have cascading effects across the software supply chain, particularly when core packages serve as the foundation to broader ecosystems.

This paper examines OSS project governance dynamics through an industry case study of \texttt{data.table}, a widely used R package for high-performance, in-memory data manipulation~\cite{dowle2022datatable}. Extensively deployed in data-driven workflows across both academia and enterprise environments, \texttt{data.table} is integral to many analytics pipelines in production settings. Despite this industrial relevance, the project's governance was initially informal and centered around a single lead maintainer---a setup that eventually struggled to support its scale and adoption. Leading up to 2023, the project faced growing challenges: a stagnating release cycle, an increasing backlog of pull requests and issues, long delays in contribution reviews, and difficulties onboarding new developers. These symptoms reflected deeper structural problems, including an absence of clearly distributed responsibilities and formal decision-making processes, which ultimately threatened the project's sustainability.

In response, the \texttt{data.table} community collaboratively began a structured transition process toward a community-driven governance model in late 2023. This effort was not top-down but emerged from open discussions within the community. The idea was brought as an issue in the issue tracker,\footnote{https://github.com/Rdatatable/data.table/issues/5676} in which contributors articulated the need for clearer roles, shared responsibilities, and a governance framework aligned with the project’s growth. The governance model was developed reflecting a collective understanding of the project's needs and a commitment to sustainable collaboration, seeking to align project management with its increasing industrial demands and community growth.

In this paper, we report an analysis of the impact of this governance reform. We used a mixed-methods approach based on a questionnaire (answered by 23 active contributors) and data mining of the project's development activity. We examine whether the new model improved transparency, participation, issue resolution, and contributor retention. Our findings reveal key benefits and remaining challenges in OSS governance and offer practical insights for other projects seeking to improve their sustainability and scalability. 

This case study highlights the potential of OSS governance in improving project health. As organizations increasingly rely on open source components, understanding how to cultivate contributor engagement, distribute responsibilities effectively, and resolve governance bottlenecks becomes not just a matter of project sustainability but a matter of software supply chain resilience. Our findings offer insights for maintainers, foundations, and companies aiming to build scalable and inclusive governance structures across OSS communities.

\section{\texttt{data.table} Project}
\texttt{data.table} is a high-performance R package developed to address the limitations of base R in efficiently manipulating large-scale, in-memory data. Since its release in 2006, it has matured into a core component of the R data science ecosystem, supporting complex operations such as fast joins, aggregations, reshaping, and parallel processing using concise, expressive syntax. The package integrates C-level optimizations and novel algorithms (e.g., radix sort) that deliver state-of-the-art performance in both computation time and memory usage. These features have made \texttt{data.table} a tool of choice in domains requiring intensive data workflows, including biomedical analysis, finance, public health, and insurance.

The impact of \texttt{data.table} is reflected in its widespread adoption. As of May 2025, over 1,000 R packages listed on CRAN directly depend on it, and it has been downloaded more than 65 million times, averaging over 1 million downloads per month~\cite{datacampstats,cranpkg}. These numbers place it among the top 40 most-downloaded R packages and one of the most central components of the ecosystem~\cite{dsmeta2025}. The package plays a foundational role in analytical pipelines used across both scientific and enterprise environments, and continues to be actively maintained and extended with recent releases introducing enhancements to row-wise operations and grouping functionalities~\cite{datatableNews}.

However, the expansion of \texttt{data.table} introduced significant challenges. The project was governed informally and relied heavily on a single core maintainer, resulting in scalability issues as contributions increased. As the user and contributor base expanded globally, the volume of issues and pull requests outpaced the capacity of a single reviewer. While the total backlog included many long-standing and low-priority items, we observed that many newly submitted issues and pull requests did not receive timely reviews or follow-up activity. For instance, before the governance changes, it was common for more than half of new contributions to remain without comments, triaging labels, or assignees for weeks---indicating insufficient review bandwidth relative to incoming activity. This is further illustrated in the backlog and pull request resolution metrics we present in Figures~\ref{fig:pull_request_backlog_count} and~\ref{fig:pull_request_mean_age}. Contributors often faced uncertainty around the contribution process, and new participants encountered barriers to onboarding due to sparse documentation and a lack of mentorship.

These governance and coordination limitations constrained the project's ability to maintain its performance standards and slowed the delivery of enhancements expected by both users and downstream package developers. Without a formal structure, prioritization was inconsistent, responsiveness varied, and community engagement suffered.

The next subsection details how the \texttt{data.table} community approached these challenges and restructured its governance model in response to the project's evolution and increased demands.

\subsection{\texttt{data.table} Governance Reform Process}

The effort to address governance challenges in \texttt{data.table} emerged within the community. As contributor activity increased and issues began accumulating, core members and new contributors recognized that relying on a single maintainer was no longer sustainable. 
At this point, the community underwent multiple concurrent changes to improve the project's health and sustainability. This included NSF-funded outreach activities, translation efforts, community-building activities, and mentorship to rebuild the community. 

The efforts to improve the project's health started with a reform in the governance model. The process was conducted in a public discussion initiated in GitHub issue \#5676\footnote{\url{https://github.com/Rdatatable/data.table/issues/5676}}, where participants raised concerns about contribution delays, undefined roles, and the lack of transparent processes. Rather than implementing top-down changes, the project embraced an open, collaborative approach to governance reform. The discussion was inclusive and iterative, with community members proposing and refining ideas for roles, responsibilities, and decision-making structures. A key motivation behind this initiative was to establish a governance model that could support both the scale of \texttt{data.table}'s usage and the diversity of its contributor base.

The final governance model introduced a tiered structure with clearly defined roles:
\begin{itemize}
    \item \textbf{Contributors} participate by submitting issues, pull requests, or translations.
    \item \textbf{Project Members} have sustained engagement and are granted permissions for creating branches in the central repository.
    \item \textbf{Reviewers} are trusted to provide technical feedback and approve changes.
    \item \textbf{Committers} have merge rights.
    \item \textbf{CRAN Maintainer} is responsible for package releases.
\end{itemize}

Beyond role definitions, the governance model outlines the process for submitting and discussing proposals, evaluating consensus, and triggering formal voting. It defines timelines and expectations for responsiveness, clarifies escalation paths for resolving conflicts, and includes mechanisms for dissent. Promotion criteria are publicly documented and tied to merit and engagement over time. The governance text also emphasizes inclusiveness, recommending multilingual documentation and community mentorship practices.

The complete governance specification is available in the project repository.\footnote{\url{https://github.com/Rdatatable/data.table/blob/master/GOVERNANCE.md}} This structured governance framework was discussed from August 2023 to November 2023 and officially adopted in December 2023. In the following sections, we evaluate its impact.

\section{Evaluation Design and Data Sources}

To evaluate the impact of the governance reform on the \texttt{data.table} project, we employed a mixed-methods approach combining contributor survey responses with repository mining techniques. This dual strategy allowed us to capture subjective perceptions from the community and objective indicators of project activity and contributor behavior over time.

\subsection{Contributor Survey}

We designed a questionnaire to capture contributor perceptions about the effectiveness of the newly adopted governance model. The instrument included closed-ended items aligned with governance-related constructs---such as transparency, decision-making clarity, contributor roles, inclusion, fairness, and overall satisfaction. These were informed by established research on OSS community health and governance~\cite{goggins2021making, trinkenreich2023belong, steinmacher2018barriers} and on templates and guidelines available in the grey literature~\cite{neary2020governance,ospo2023ggi,osiGovernance,apacheGovernance}. The questions were refined through discussion among researchers and practitioners with expertise in OSS ecosystems. The 17 closed-ended questions are listed in Table~\ref{tab:surveyQuestions} and were presented using a five-point Likert scale ranging from ``Strongly Disagree'' to ``Strongly Agree,'' with a neutral midpoint.

\begin{table}[ht]
    \centering
    \caption{Survey Questions}
    \label{tab:surveyQuestions}
    \footnotesize
    \begin{tabular}{p{0.35cm}|p{6.5cm}}
        \hline
        \textbf{Id} & \textbf{Description} \\\hline
        \multicolumn{2}{|c|}{\textbf{Governance 
Transparency}} \\\hline
        Q1& The new governance model has made \texttt{data.table}'s decision-making process more transparent. \\\hline
        Q2 & Based on the new governance model, I feel that it is easy to understand how decisions are made in \texttt{data.table} project. \\\hline
        Q3 & The new governance model clearly communicates \texttt{data.table}'s goals and vision. \\\hline
        
        \multicolumn{2}{|c|}{\textbf{Community Involvement}} \\\hline
        Q4 & The governance model encourages contributors to participate actively in \texttt{data.table} activities. \\\hline
        Q5 & The new governance model makes me feel more valued as a contributor. \\\hline

        \multicolumn{2}{|c|}{\textbf{Roles and Responsibilities}} \\\hline
        Q6 & The new governance model clearly defines the existing roles within \texttt{data.table}. \\\hline 
        Q7 & The new governance model makes me understand my responsibilities better now than before. \\\hline
        Q8 & The collaboration within the community has improved after the new governance model. \\\hline

        \multicolumn{2}{|c|}{\textbf{Conflict Handling}} \\\hline
        Q9 & The governance model implementation improved fair decision-making. \\\hline
        Q10 & Conflicts within the community are now handled more effectively. \\\hline
        Q11 & I feel that my opinions are considered in important project decisions. \\\hline

        \multicolumn{2}{|c|}{\textbf{Code of Conduct and Community Culture}} \\\hline
        Q12 & The governance model has improved the overall community behavior. \\\hline
        Q13 & The code of conduct helps to incentivize fairness. \\\hline
        Q14 & The project is now a more inclusive and respectful environment. \\\hline

        \multicolumn{2}{|c|}{\textbf{Engagement and Project Health}} \\\hline
        Q15 & The governance model has increased my engagement with the project. \\\hline
        Q16 & The community seems more active since the governance model was introduced. \\\hline

        \multicolumn{2}{|c|}{\textbf{General Feedback}} \\\hline
        Q17 & I am satisfied with the project's governance model. \\\hline
    \end{tabular}
\end{table} 

The survey was distributed to the \texttt{data.table} community via the GitHub issue tracker,\footnote{\url{https://github.com/Rdatatable/data.table/issues/6715}}. The survey was available between January and February 2025 and received 25 responses. We removed responses with blank values in at least one question on the survey (n=8). Post filtering, our dataset includes 17 responses for analysis, resulting in a 44.73\% response rate (considering that 38 unique contributors were involved in activities in the project GitHub between February 2024 and February 2025). The response rate can be considered high compared to recent Software Engineering surveys, which range between 13\% and 44\%~\cite{zhang2019companies}, \cite{dilhara2022discovering}. The demographics of the respondents are presented in Table \ref{tab:demographics}.

\begin{table}[ht]
    \caption{Demographics of Respondents}
    \label{tab:demographics}
    \centering
    \footnotesize
    \begin{tabular}{l|r}
        \hline
        \textbf{Attribute} & \textbf{Count} \\\hline
        
        \multicolumn{2}{|c|}{\textbf{Gender}} \\\hline
        Men & 15\\
        Women & 1\\ 
        Non-Binary/Third Gender & 1 \\\hline
        
        \multicolumn{2}{|c|}{\textbf{Experience in Years}} \\\hline
        Less than 1 year & 8 \\
        1 to 3 years & 1 \\
        3 to 7 years & 5 \\
        More than 7 years & 3\\\hline
        
        \multicolumn{2}{|c|}{\textbf{Role}} \\\hline
        Code Development & 12 \\
        Code Review & 3 \\
        Bug Triaging & 4 \\
        Translation & 4 \\
        Documentation & 2 \\
        Community Building & 1 \\
        Advocacy and Evangelism & 2 \\
        Mentorship & 1 \\
        User Support & 2 \\ \hline
    \end{tabular}
\end{table}

We added the questions used in the study in the replication package \cite{arantes2025datatable_replication}. For confidentiality, the raw data with demographic question answers could not be shared to avoid participants' identification. 

\subsection{Repository Mining}

In parallel, we collect data from the \texttt{data.table} GitHub repository history to extract quantitative indicators of community activity before and after the governance changes. The metrics used to analyze community behavior were discussed during meetings between October 2024 and November 2024 in a group of researchers experienced in OSS, based on CHAOSS metrics~\cite{CHAOSS}. The selected metrics provide a general development process perspective, including pull request creation and review, and issue creation and support. These metrics were chosen because they reflect the activities in the two main channels of the project (issue tracker and pull requests) and reflect the influx and retention of contributors, key to measuring the health and sustainability of the project. 

Thus, our analysis focused on the following core metrics, each supported by recent empirical studies in software engineering and open source research:

\begin{itemize}
    \item \textbf{Pull Request Age}: the average time (in days) between opening and closing for pull requests created in each month, reflecting review/merge throughput \cite{hasan2023understanding}, \cite{khatoonabadi2024predicting}.
    \item \textbf{Backlog Size}: the number of unresolved (open) pull requests and issues, indicating responsiveness and review capacity \cite{varanasi2024measuring}.
    \item \textbf{Active Contributors}: the number of unique contributors opening pull requests per month \cite{li2022exploring}, \cite{stuanciulescu2022code}.
    \item \textbf{New Contributors}: contributors opening their first PR each month, as an indicator of engagement/recruitment \cite{stuanciulescu2022code,fang2023matching}.
    \item \textbf{Retained Contributors}: contributors who submitted PRs across multiple months, indicating retention \cite{stuanciulescu2022code,fang2023matching}.
    \item \textbf{Activity Ratio}: the proportion of PRs that received comments or review labels within two weeks~\cite{hasan2023understanding,khatoonabadi2024predicting}.
\end{itemize}

Trend lines were applied to support the analysis of temporal changes, enabling us to detect behavior shifts. The data was selected across a 25-month window, partitioned into two 12-month periods: pre-governance (November 2022 – November 2023) and post-governance (January 2024 – January 2025). We excluded December 2023 as this was the month when the new governance model was implemented and may bring noisy data. We added the mined data used in the study in the replication package~\cite{arantes2025datatable_replication} as well. 

\subsection{Trend Comparison Analysis}

To supplement the descriptive repository mining results with formal temporal analysis, we implemented a custom method to compare pre- and post-intervention trends in key metrics. Specifically, we compute and contrast linear trends across two time segments using monthly aggregated data. We used 12 months before/after the governance model, as following: November 2022–November 2023 (pre-governance) and January 2024–January 2025 (post-governance). December 2023 was excluded as a transition month.

The method accepts a time series of the metric analyzed(e.g., pull request closure counts, issues open) and performs the following steps: (1) filters the dataset to the defined pre- and post-intervention time windows; (2) applies ordinary least squares (OLS) regression separately to each segment to estimate the slope, intercept, and model fit (R$^2$); and (3) tests the statistical significance of the difference in slopes using an interaction model of the form 

\texttt{metric \(\sim \) time + segment + time:segment}

where \textit{time} is the time index within each segment (e.g., 0, 1, 2, ..., n);  \textit{segment} is a binary variable indicating whether a data point belongs to the pre- (0) or post- (1) governance period; and \textit{time:segment} is the interaction between time and segment --- it captures how the slope (i.e., rate of change over time) differs between the two periods.

By analyzing the difference in the slope, we could to assess whether the governance reform led to a statistically significant shift in the trajectory of a given metric. We applied this method across metrics such as pull request resolution rate, contributor activity, and issue backlog to quantify the governance model's impact on project dynamics. 

\section{Survey Results}
The 17 closed-ended questions are listed in Table~\ref{tab:surveyQuestions}. The survey responses provided insight into contributor perceptions regarding the impact of the governance model. Below, we summarize the results by grouping the questions into thematic governance dimensions that reflect the constructs captured in the survey design.

\begin{figure*}[ht!]
    \centering
    \includegraphics[width=14cm]{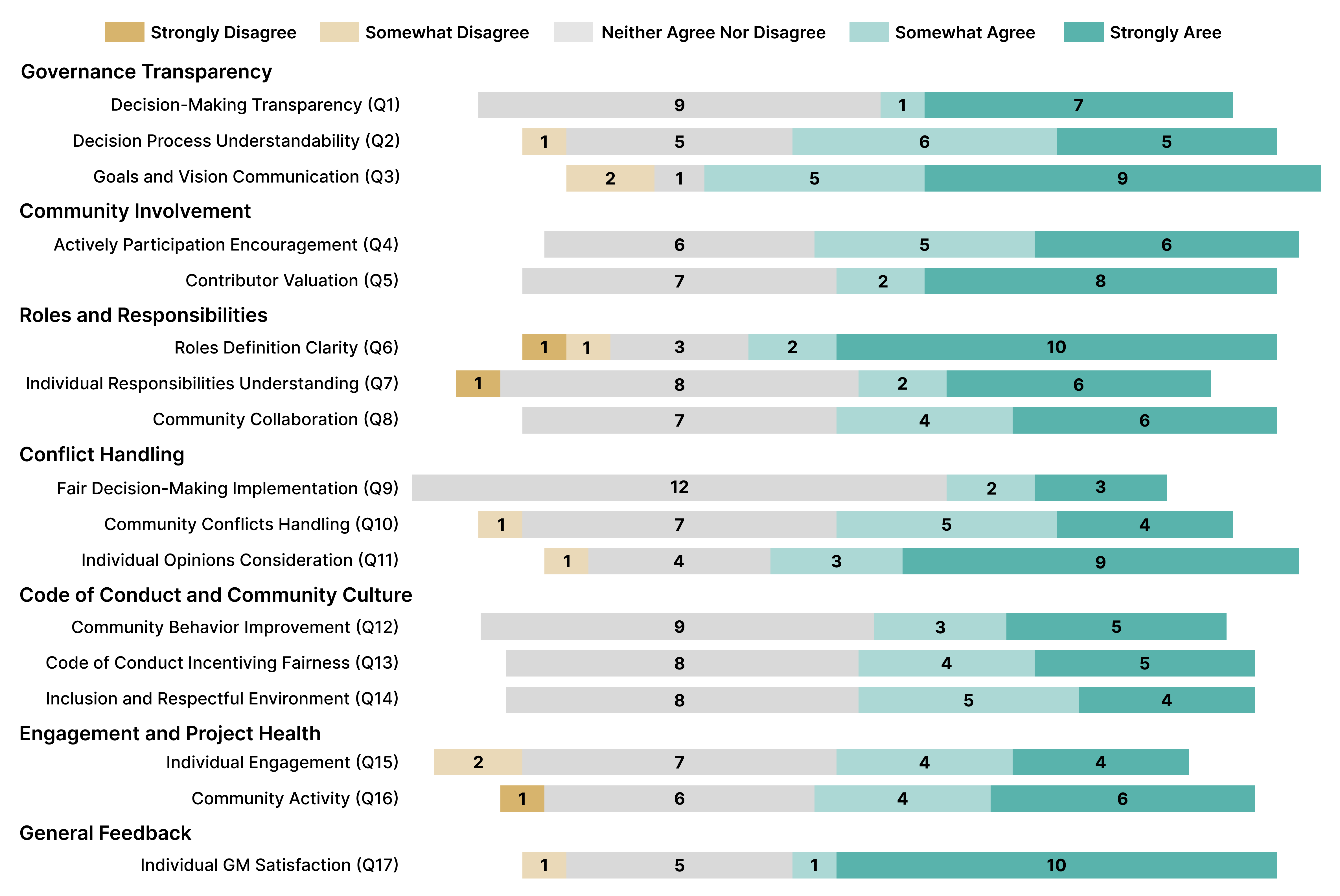} % Adjust width as needed
    \caption{Survey Results}
    \label{fig:likertChartAll}
\end{figure*}

We have grouped the survey questions into categories reflecting key aspects of OSS governance and community dynamics. Each category represents a specific dimension, including transparency, community involvement, roles and responsibilities, decision-making processes, community culture, and overall project health. %We performed a Confirmatory Factor Analysis (CFA), which showed that most of the items presented good/excellent loading scores ($>0.7$) in almost all cases, even with a rather small sample size. Exceptions are Q13 and Q15, which loaded weakly. The results of CFA are presented in Table~\ref{tab:CFA}.
The following sections present the survey findings for each group.

\subsection{Transparency and Decision-Making Clarity}

Participants expressed generally favorable perceptions regarding governance transparency introduced by the new model. Almost half of the responses indicated agreement or strong agreement that decision-making in the project has become more transparent (Q1), with no one disagreeing. Similarly, respondents tended to agree that the decision-making process is now easier to understand (Q2). In terms of communication about goals and vision, responses were even more positive, with a notable concentration of participants selecting ``strongly agree.'' However, a few participants expressed disagreement regarding decisions (1 participant) and the communication of goals and vision of the project (2 participants). These respondents had over three years of involvement with the project, suggesting that while the model is positively received overall, its impact may not be equally perceived by long-term contributors. This highlights an opportunity for improvements in the transparency strategies.

\subsection{Community Involvement}

Community involvement in OSS projects refers to the active participation of contributors in multiple activities, such as code contributions, issue discussions, documentation improvements, and decision-making processes. 
According to our respondents, the new governance model appears to have had a positive effect on encouraging this involvement. Most participants agreed or strongly agreed that the model encourages greater participation. Responses also indicated that individual contributions are valued under the new model. The absence of disagreement and the strong concentration of positive responses suggest that community members generally feel more encouraged and appreciated. This indicates that the governance changes have had a constructive impact on fostering a more participatory community environment.

\subsection{Roles and Responsibilities}
For OSS projects, defining roles and responsibilities is essential for efficient collaboration and clear decision-making. A well-structured governance model helps contributors understand their roles and fosters an organized, productive development process.

Participants' responses to this set of questions were positive, in general. Most respondents agreed or strongly agreed that the governance model provided a clearer role definition (Q6). When asked about their understanding of individual responsibilities (Q7), responses trended to a positive end, but half of the respondents were neutral about it. One can notice that Q6 and Q7 received negative responses (by two respondents with more than 3 years in the project), indicating that some contributors do not perceive a significant improvement in structural clarity. Finally, collaboration within the community (Q8) was also viewed positively. These responses suggest that, while the governance model has enhanced the clarity of the structure by proposing explicit roles, its impact was not perceived by some more experienced contributors. This may reflect a mismatch between formal role documentation and informal team dynamics, or expectations built over time that are not being met under the new structure.

\subsection{Conflict Handling}
Effective conflict resolution is critical in any open source governance model. A transparent and inclusive decision-making process ensures that contributors feel heard and maintain a collaborative and productive environment.
According to our answers, feedback on the improvement on decision-making fairness (Q9) was largely neutral, with most participants selecting the midpoint of the scale (70\%), but no respondents indicating disagreement. In contrast, responses to Q10, which focused on the resolution of conflicts, were more divided: while most contributors agreed with improvements (53\%), a portion remained neutral, and one disagreed. Similarly, Q11, concerning the contributors’ opinions during decision-making, received mainly positive responses, though with one disagreement. Notably, the few negative responses in Q10 and Q11 came from contributors with more than three years of involvement. This suggests that there is still room to improve the conflict handling in the project, or providing more information about this to the community is required.

\subsection{Code of Conduct and Community Culture}
A well-defined code of conduct and a strong community culture are essential for fostering a respectful, inclusive, and collaborative environment in open source projects. A governance model should promote positive interaction and enforce fair community standards, creating a welcoming space for all contributors.
In this category, responses were generally divided between positive and neutral perspectives across all questions. Contributors tended to agree that the governance model has improved overall community behavior (Q12). They also agreed that the code of conduct plays an important role in promoting fairness (Q13). Similarly, most respondents viewed the project as now being a more inclusive and respectful environment (Q14). While some participants selected neutral responses, notably, no one disagreed with any of the statements in this dimension. The absence of negative sentiment suggests that the governance model is seen as effective in fostering a positive and respectful community culture, making this one of the most favorably perceived aspects of the recent changes.

\subsection{Engagement and Project Health}
Engagement and overall project health are key indicators of a thriving open source community. As noted by Trinkenreich \textit{et al.}~\cite{trinkenreich2023belong}, ``the sustainability and long-term survival of Open Source Software (OSS) projects depend not only on attracting but, more crucially, retaining motivated developers.'' 

Responses for the items under this category reflected moderately positive perceptions. Many participants agreed that the governance model increased their engagement with the project (Q15), though some opted for neutral responses. Perceptions of broader community activity (Q16) were slightly more favorable, with several participants indicating that the project appears more active since the governance model was introduced. The few negative responses in both questions came, once again, from contributors with more than three years in the project. This suggests that the governance model reform may not have fully translated into renewed personal engagement for every experienced contributor, possibly due to differences in expectations or historical engagement patterns.

\section{Data Mining Results}

To complement survey insights, we analyzed project activity metrics extracted from the \texttt{data.table} GitHub repository. The figures below illustrate longitudinal trends in pull request resolution, contributor activity, and backlog dynamics, comparing behavior before and after the implementation of the governance model in December 2023.

\subsection{Pull Request Resolution Time}

\begin{figure}[tb]
    \centering
    \includegraphics[width=.47\textwidth]{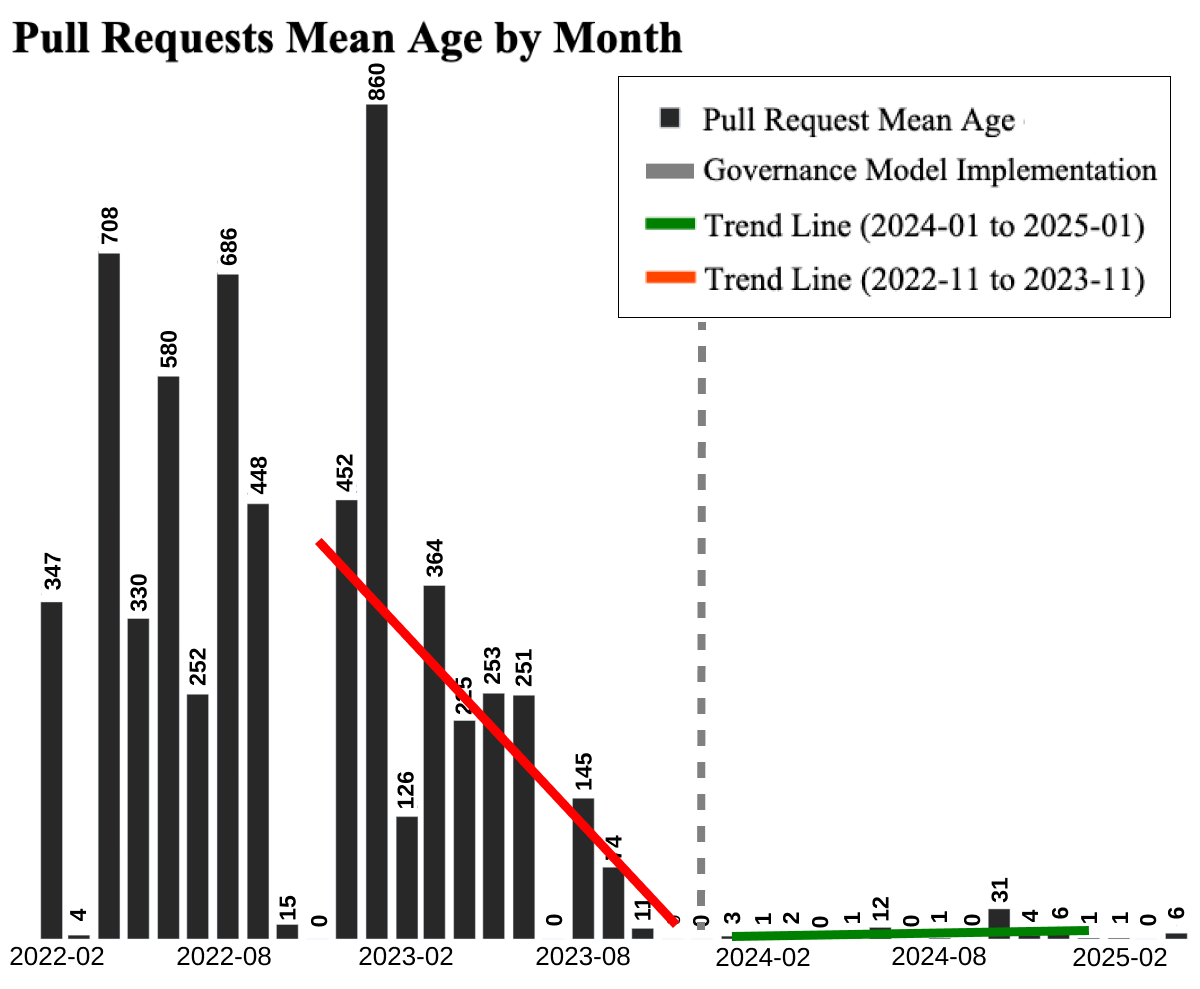}
    \caption{Pull Request Average Resolution time}
    \label{fig:pull_request_mean_age}
\end{figure}

Figure \ref{fig:pull_request_mean_age} illustrates the average resolution time for pull requests created in each month, from February 2022 through October 2024. Initially, the data reveals that before the implementation of the new governance model in early 2024, the average age of pull requests peaked at approximately 860 days, indicating a significant delay in addressing contributions. Notably, this period is marked by a gradual accumulation of unresolved requests, suggesting potential bottlenecks in the review process.

Following the governance model implementation, a dramatic turnaround is observed. Within a few months, the average resolution time converges to as low as 4 days for the pull requests opened after the new governance model. This shows a rapid improvement in the community's responsiveness. 

The trend lines further illustrate these changes: while the 12-month trend before the reform shows a decline in PR resolution time, the consistently high average values indicate that many of those PRs were only resolved after the governance model took effect. In contrast, PRs created after the reform exhibit uniformly short resolution times, reflecting immediate and sustained attention to new contributions.

By comparing the linear regressions before/after the governance model implementation, we observed a significant decline in the pull request resolution time during the pre-reform period (slope = –56.87 days/month, p = 0.0066), followed by a plateau post-reform (slope = +0.64, p = 0.5648). The change in slope was statistically significant ($\Delta$slope = +57.52, p = 0.0053), confirming that the growing PR backlog halted and the processing times were stabilized.

One may question whether there was not enough time to assess the long-term impact since the PRs opened in this period are rather new. Although this is true, the monthly average for the post-governance period is close to zero for almost every month, showing proactivity towards the new contributions coming to the repository.

\subsection{Backlog Analysis}
\begin{figure}[tb]
    \centering
    \includegraphics[width=8.5cm]{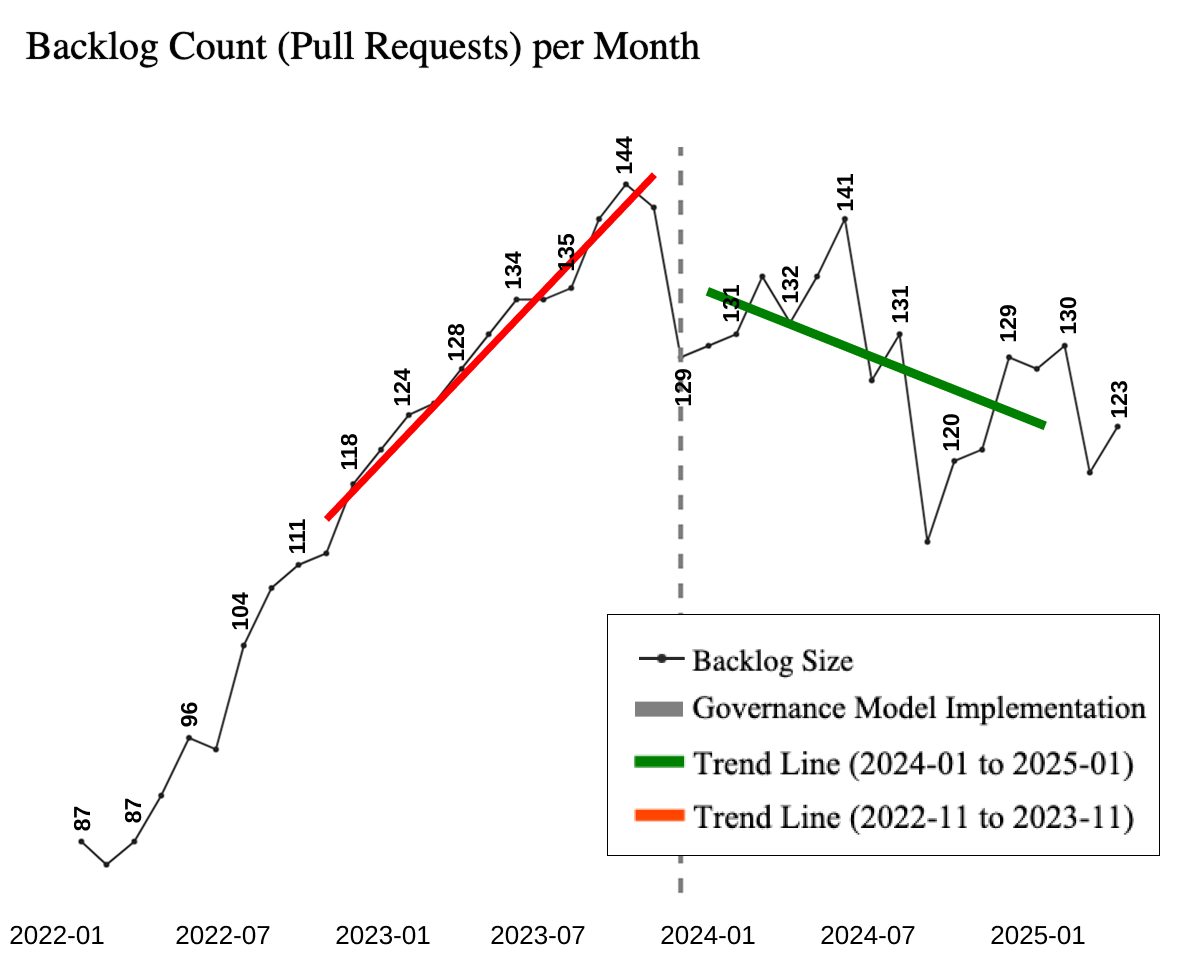} % Adjust width as needed
    \caption{Predominant Community Pull Request Activity}
    \label{fig:pull_request_backlog_count}
\end{figure}

A project backlog is calculated as the number of unresolved issues and unrevised pull requests over a specific period. A large backlog can indicate inefficiencies in addressing user needs or community disengagement. Figure~\ref{fig:pull_request_backlog_count} shows the monthly accumulated backlog for \texttt{data.table}. The pre-reform period is characterized by a persistently large (and increasing) backlog, at times exceeding 120 unresolved items. Following the adoption of the new governance model, this number steadily declined, suggesting that the community was able to resolve outstanding contributions more effectively. 

Our temporal analysis confirms this, showing that the size of the backlog increased significantly before the reform (slope = +2.46, p $<$ 0.0001) but began to decline post-reform (slope = –0.96, p = 0.0819). The difference in slopes was significant ($\Delta$slope = –3.41, p $<$ 0.0001).
This reduction reflects a stronger commitment to maintaining a manageable queue of pending contributions.

\subsection{Conclusion Rate}
The conclusion rate is defined as the difference between the number of pull requests opened and those closed each month. A high conclusion rate suggests the community is capable of addressing incoming contributions. This metric complements the understanding brought in the previous section, providing context on how the backlog increased.

\begin{figure}[tb]
    \centering
    \includegraphics[width=8.5cm]{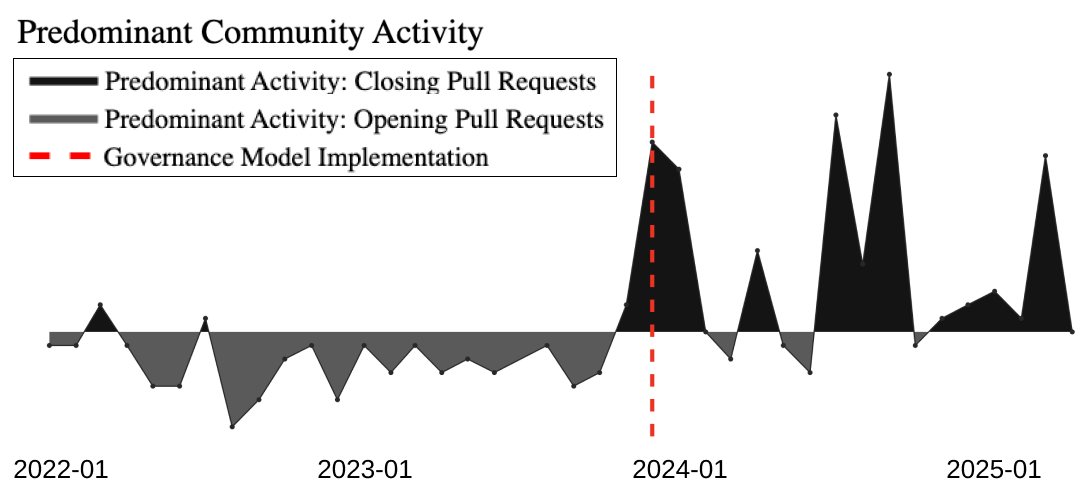} % Adjust width as needed
    \caption{Predominant Community Pull Request Activity}
    \label{fig:predominant_community_pr_activity}
\end{figure}

\begin{figure}[tb]
    \centering
    \includegraphics[width=8.5cm]{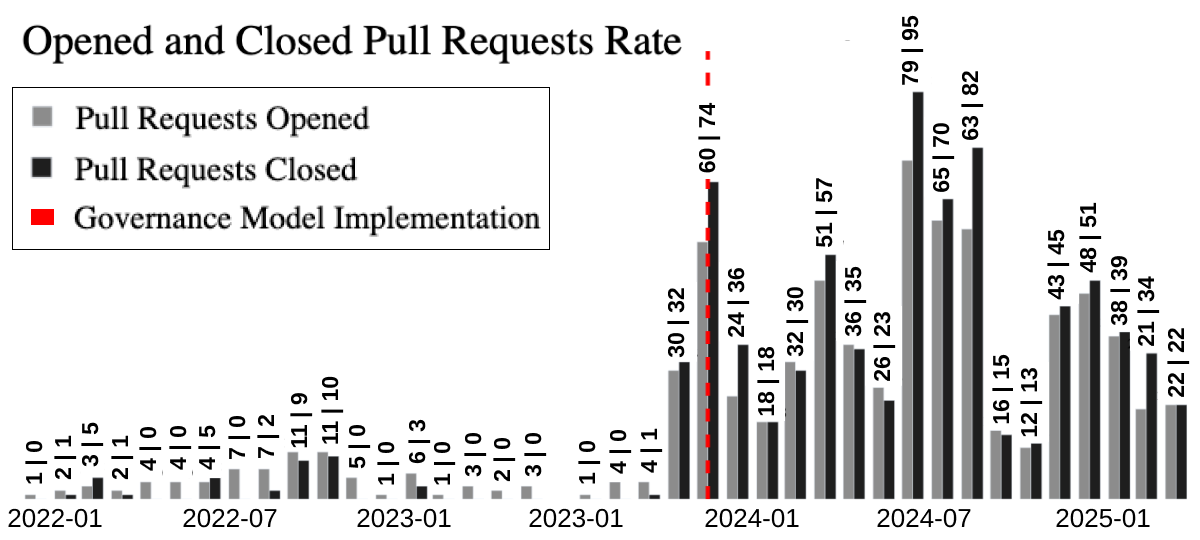} % Adjust width as needed
    \caption{Open and Closed Pull Requests By Month}
    \label{fig:opened_closed_pull_requests_rate}
\end{figure}

Figure \ref{fig:predominant_community_pr_activity} shows that before the governance model, there was a consistent trend of opening pull requests as the predominant activity (points below the line). Following the new governance model, the activity pattern shifted, emphasizing the closing of pull requests as the predominant activity. This perspective is complemented by Figure \ref{fig:opened_closed_pull_requests_rate}, which illustrates the total pull requests opened and closed per month. Both the number of opened and closed pull requests increased, slowly paying the debt from previous years (explaining the decreasing trend in the backlog --- Figure~\ref{fig:pull_request_backlog_count}).

\subsection{Community Engagement and Growth}

The open source community's health and sustainability are usually tied to the consistent activity and engagement of its members. Active participation ensures that the project evolves to meet users' needs and maintains a collaborative atmosphere. Without fostering continuous engagement, projects risk stagnation.

\begin{figure}[tb]
    \centering
    \includegraphics[width=8.5cm]{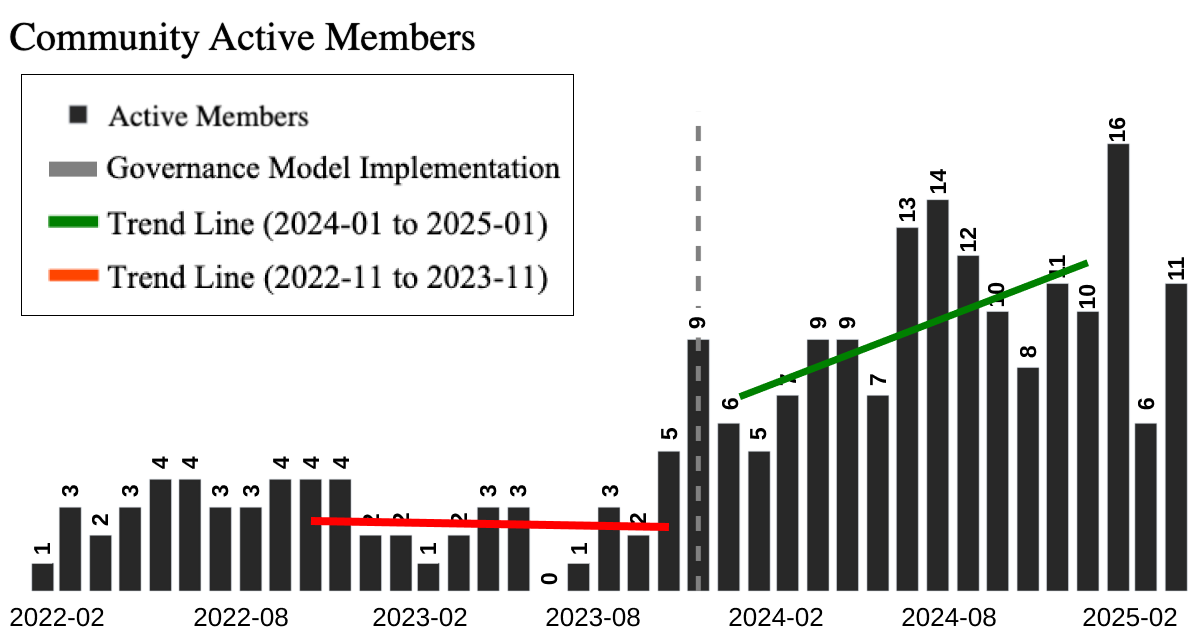}
    \caption{Active Members in Community (Pull Request Openers)}
    \label{fig:community_pr_active_members}
\end{figure}

Figure~\ref{fig:community_pr_active_members} depicts the number of unique contributors opening pull requests each month. Before the governance reform, the number of active contributors fluctuated around 3--4, with several months of near-zero engagement. In contrast, the post-reform period shows a consistent increase, peaking at 14 active contributors in one month. %This rise indicates a revitalized community and improved participation due to clearer onboarding paths and inclusive policies.

Notably, five months before the implementation, there was an entire month with zero active members, underscoring the dramatic shift in engagement after implementation. 

Temporal analysis showed that the number of monthly PR openers showed no trend before the reform (p = 0.95), but increased afterward (slope = +0.41), with a marginally significant slope difference (p = 0.0729). %This supports the interpretation that governance reforms improved contributor engagement.

\begin{figure}[tb]
    \centering
    \includegraphics[width=7.5cm]{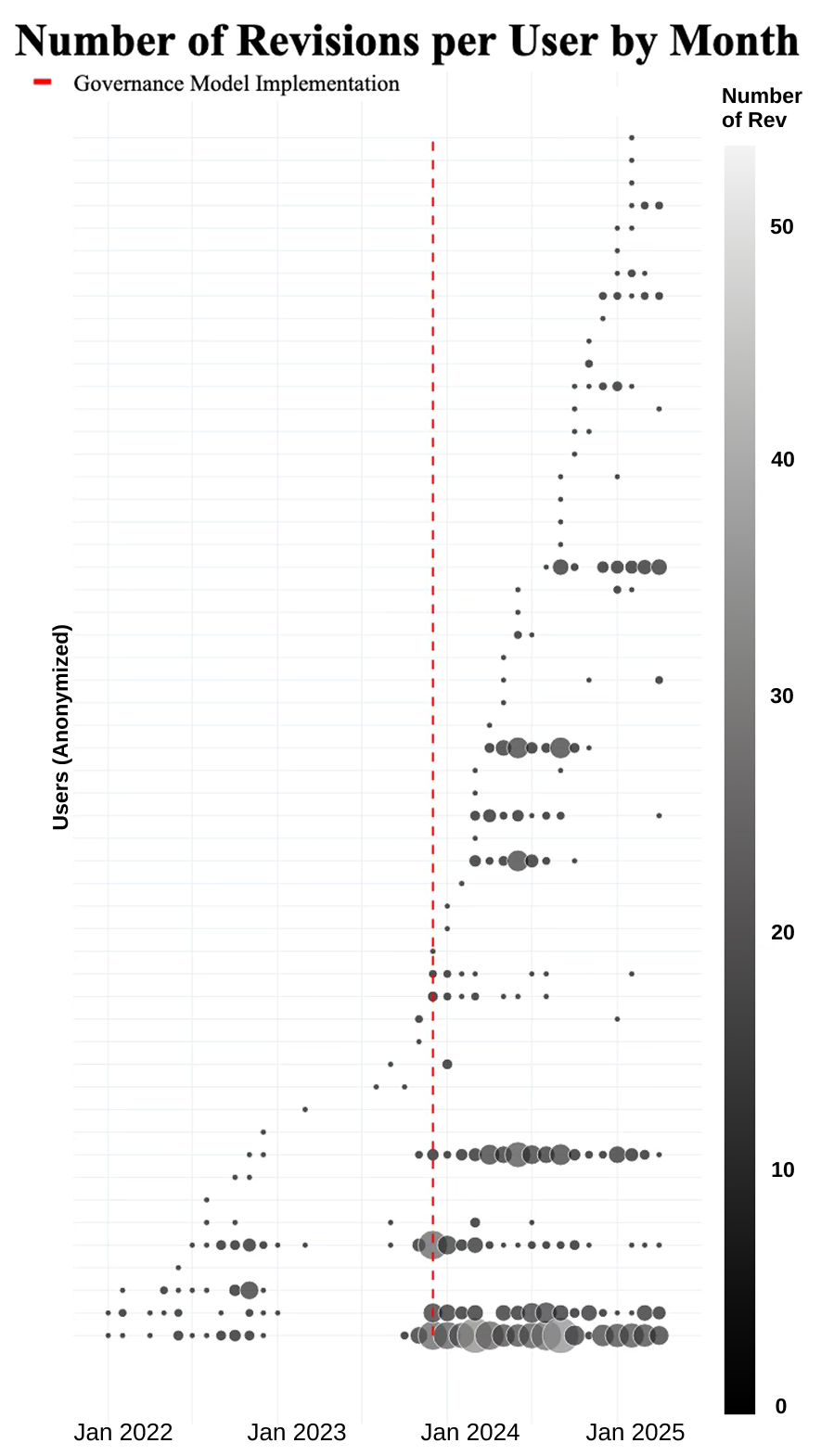}
    \caption{Pull Request Reviewed by User}
    \label{fig:pullRequestReviewedBubble}
\end{figure}

\begin{figure}[tb]
    \centering
    \includegraphics[width=7.5cm]{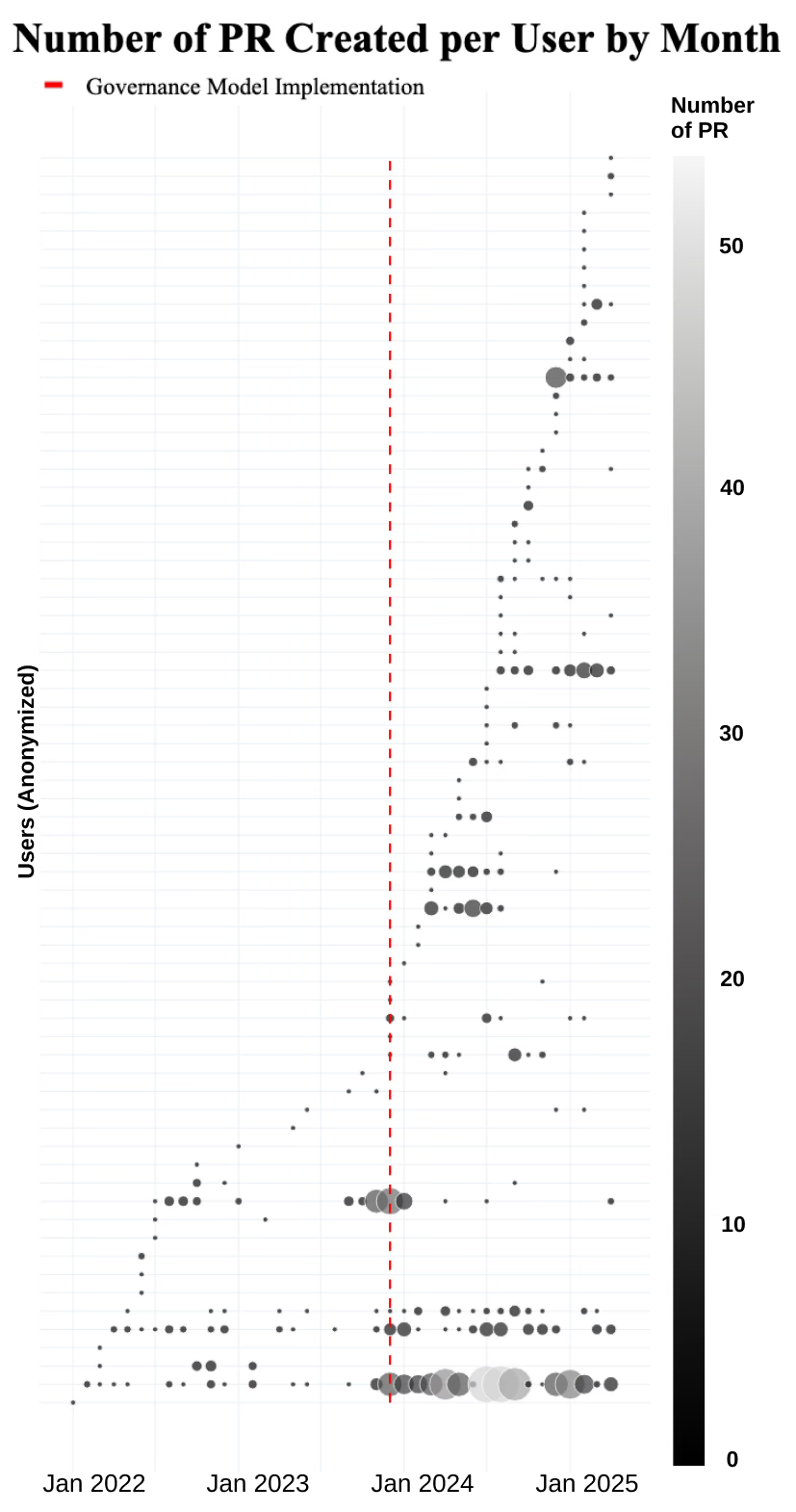}
    \caption{Pull Request Created By User}
    \label{fig:pullRequestCreatedBubble}
\end{figure}

To further understand the distribution of contributor efforts, Figures~\ref{fig:pullRequestReviewedBubble} and~\ref{fig:pullRequestCreatedBubble} present the number of pull requests reviewed and created per user by month, with each dot representing a contributor and the size indicating activity intensity.

As shown in Figure~\ref{fig:pullRequestReviewedBubble}, review activity was historically concentrated in a small number of users before governance reform, with the bottom rows (indicating few reviewers) dominating from 2022 to late 2023. Post-implementation, there is a visible diffusion of review responsibilities, with increased participation across a broader set of contributors. Notably, several new contributors began reviewing after January 2024, indicating the governance model's success in delegating responsibility and promoting reviewer onboarding.

Similarly, Figure~\ref{fig:pullRequestCreatedBubble} shows the number of pull requests submitted by users over time. Before the governance changes, contribution activity was sporadic and skewed toward a few participants. After the reform, submission activity grew more distributed and frequent, with a wider base of contributors submitting pull requests. %This pattern reflects increased openness and better onboarding pathways for contributors to engage in the development process.

\subsection{New and Retained Contributors}
Attracting, onboarding, and retaining new contributors is crucial for long-term success on open source projects~\cite{steinmacher2014attracting}. The retention rate is determined by measuring the difference between the total number of active developers and the number of newly joined developers in a specific month.

\begin{figure}[tb]
    \centering
    \includegraphics[width=8.5cm]{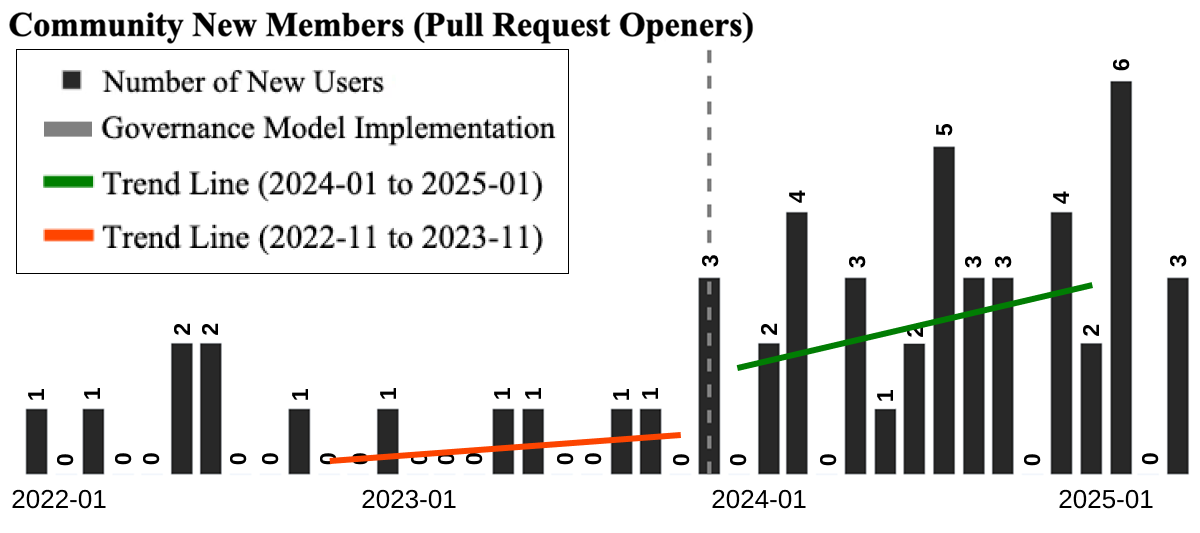}
    \caption{Community New Members (Pull Request Openers)}
    \label{fig:community_pr_new_members}
\end{figure}

In terms of attraction and onboarding new members, Figure \ref{fig:community_pr_new_members} shows that, before the governance model implementation, the number of new members stayed around 1 per month. Following the governance model implementation, the community constantly attracted around 3 developers per month, with a peak of 5 developers. Although the visual shows higher numbers, the temporal analysis of the slope of new contributors per month did not show statistically significant changes in terms of slope ($\Delta$slope = +0.055, p = 0.8820). 

\begin{figure}[tb]
    \centering
    \includegraphics[width=8.5cm]{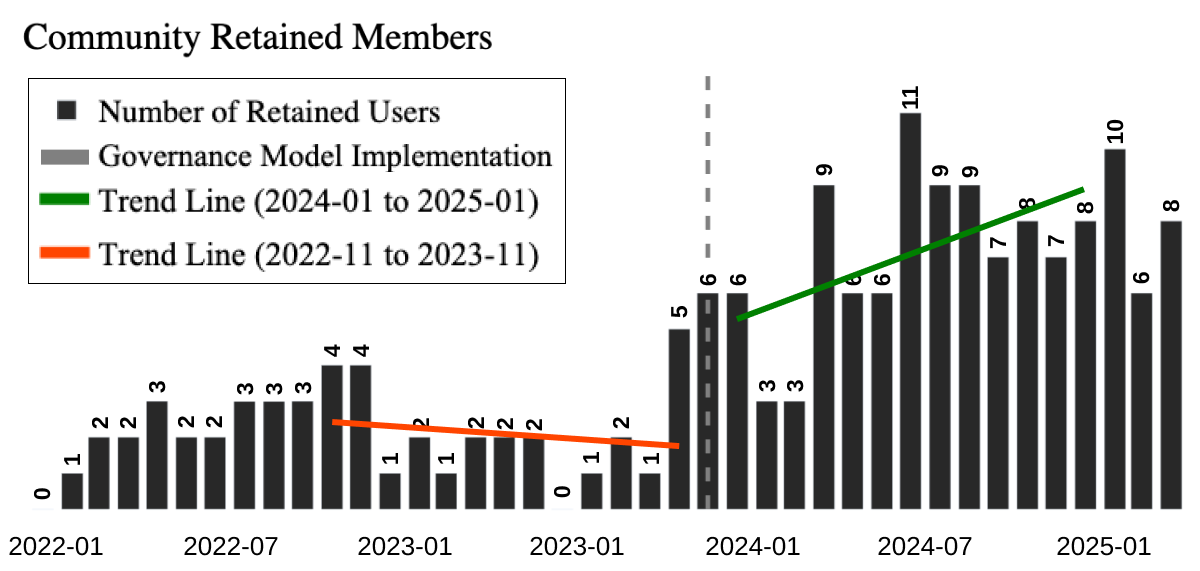}
    \caption{Community Retained Members (Pull Request Openers)}
    \label{fig:community_retained_pr_members}
\end{figure}

Retention also shows higher values after the new governance model. Figure \ref{fig:community_retained_pr_members} shows that in addition to attracting, there was a sustained growth in retained contributors, with an average of 7 contributors returning in average per month (max: 11, min: 3, median: 6.5). 

The results of the regression trend analysis show that, while the pre-reform period showed no growth (slope = –0.038, p = 0.7519), the post-reform period exhibited a positive trend (slope = +0.30, p = 0.0777). However, the slope difference was marginally significant (p = 0.1058), possibly due to the short post-intervention window.

With more new members joining and existing contributors staying engaged, the community benefits from a richer diversity of ideas and a more resilient support network~\cite{vasilescu2015gender}. The balance between recruiting new talent and retaining experienced contributors is a key indicator of a healthy, sustainable open source community.

\section{Discussion}

The results from the questionnaire and repository mining converge to illustrate a significant transformation in the \texttt{data.table} ecosystem following the introduction of the new governance model. By integrating qualitative feedback with quantitative metrics, we draw a comprehensive picture of how governance changes have influenced operational efficiency, community dynamics, and long-term sustainability.

\subsection{Governance as a Catalyst for Operational Turnaround}

Quantitative evidence revealed a dramatic reduction in pull request resolution time---from peaks above 700 days to just a few days within months of governance implementation. This improvement is especially important in industrial contexts, where responsiveness and stability are key to ensuring that OSS dependencies remain viable. Moreover, the survey confirms that contributors perceived a parallel increase in transparency and clarity of roles, suggesting that governance changes were not only operationally effective but also visible and credible to the community.

\subsection{Engagement, Recruitment, and Retention Gains}

The sustained increase in both new and retained contributors post-reform is one of the most compelling findings. Contributor retention is a long-standing challenge in OSS communities~\cite{trinkenreich2023belong}, and the ability to increase not just recruitment but continued engagement suggests that the governance model addressed systemic obstacles to participation. Features such as role progression, shared review duties, and documentation improvements likely contributed to this effect. Survey responses indicating a stronger sense of being valued support this interpretation.

\subsection{Mixed Results on Fairness and Dispute Resolution}

While overall perceptions were positive, not all dimensions improved uniformly. Responses to questions around fairness, decision-making, inclusivity, and conflict resolution were more muted or neutral. This may reflect that governance structures---even well-defined ones---do not automatically lead to perceived fairness unless conflict-handling processes are enacted visibly and consistently. %The results imply that building trust in dispute resolution mechanisms may require both more communication and additional governance maturity over time.

\subsection{Integrated Insights and Implications for Sustainability}
When synthesizing the survey and data mining findings, it becomes evident that the new governance model has catalyzed positive changes within the \texttt{data.table} ecosystem. Enhanced transparency and clearer role definitions may have improved community engagement, pull request resolution time, and effective management of backlogs. 

Moreover, the significant increases in both recruitment and retention rates point to a more dynamic and sustainable community structure. However, the mixed feedback on conflict resolution and fairness indicates that the governance model, while successful in many areas, still faces challenges in fully addressing all aspects of community dynamics. 

\subsection{Broader Implications for Industry-Grade OSS}

The \texttt{data.table} case underscores that even well-established OSS projects with strong technical foundations can falter without scalable governance. Its revitalization illustrates that technical excellence alone is insufficient: sustained health in OSS ecosystems also depends on how responsibilities are distributed, how contributors are supported, and how decisions are made. For organizations that rely on OSS infrastructure, these results reinforce the value of investing not just in code, but in community processes and contributor experience.

Importantly, the structured governance adopted here was achieved through consensus-driven, community-led reform. This highlights the feasibility of adapting governance models in community-based OSS settings and suggests that similar interventions could be both desirable and achievable in other widely used OSS projects.

\subsection{Toward Evidence-Based Governance in OSS}

Finally, this study contributes to the growing call for empirical grounding in OSS governance research~\cite{goggins2021making}. Rather than treating governance as a static artifact, this work demonstrates how governance reform can be adjusted according to participation challenges. Future work should investigate how governance evolution unfolds across different organizational structures and whether lessons from \texttt{data.table} generalize to other contexts.

\section{Threats to Validity}

%Although this is an industry-oriented paper, the empirical nature of our assessment makes it necessary to acknowledge the threats to validity of this paper.

\noindent\textbf{Construct Validity.} Since there was no validated instrument to assess the governance model, we designed our items based on prior literature and best practices in OSS governance research and grey literature. Moreover, the items rely on self-reported perceptions, which may be influenced by individual expectations, social desirability bias, or limited visibility into project-wide processes. Although Likert-type scales enable consistent interpretation, responses may not fully capture contributors' experiences. However, the high completion rate, alignment between qualitative and quantitative findings, and consistency in responses across participant profiles suggest that the survey instrument was effective in capturing relevant perceptions.

\noindent\textbf{Internal Validity.} Causal attribution is a challenge in real-world OSS settings. While temporal alignment between governance changes and improved metrics is evident, we cannot definitively rule out alternative explanations (e.g., seasonal activity shifts, unrelated external factors, or community momentum). It is also important to reinforce that the community underwent multiple concurrent changes during the same period (NSF grant to support outreach, translation, and community-building activities). These efforts likely reinforced the effects of governance reform and should be considered as complementary interventions. 
However, the sharp, immediate, and sustained nature of the observed improvements---coupled with alignment across both subjective and repository data---suggests that the governance changes played a catalytic role. 

\noindent\textbf{External Validity}
This is a case study of \texttt{data.table}, thus, the generalizability of our findings is limited. While \texttt{data.table} is widely used and industrially relevant, governance dynamics may vary across projects with different sizes, community structures, or funding models. Replication across additional projects and ecosystems would be needed to establish broader applicability. However, \texttt{data.table} represents a particularly informative case due to its combination of large user base, long-term maintenance history, and practical deployment, making it a strong baseline for understanding governance reform in mature, high-impact OSS projects.

\noindent\textbf{Conclusion Validity.}
Our repository analysis uses standard activity metrics (e.g., pull request age, backlog size, contributor retention), but these proxies do not capture all relevant aspects of governance efficacy. In addition, visual trend interpretation, while useful, is inherently descriptive and not grounded in formal statistical testing. Our conclusions are therefore best understood as observational and indicative rather than definitive. However, the consistency between behavioral trends and survey responses across multiple governance dimensions lends robustness to the interpretation that the observed improvements were driven by structural governance changes.

Despite these limitations, the combination of community-grounded survey data and empirical behavioral indicators provides a robust foundation for understanding the effects of governance reform. Future work should build on this with longitudinal and cross-project designs to further validate and extend these insights.

\section{Conclusions}
This paper presents the results of a governance reform undertaken by the \texttt{data.table} community in response to concrete organizational and operational challenges. Before the governance change, the project faced increasing coordination bottlenecks, growing backlog, low contributor retention, and an overreliance on a single maintainer---threatening project sustainability despite strong technical foundations and widespread practical use. Motivated by these concerns, the community collaboratively developed and adopted a structured governance model, formalizing roles, decision-making processes, and contributor pathways. This paper documents the potential effects of that transition using a mixed-methods approach that integrates contributor perceptions with objective repository metrics.

We observed substantial improvements in responsiveness, contributor engagement, and project momentum following the implementation of the governance model. Contributors reported greater clarity in project direction and roles, and repository mining revealed increases in contributor retention, review throughput, and activity responsiveness. While perceptions around fairness and conflict resolution remained more neutral, the overall effect of the governance reform has been transformative.

Our findings highlight that structured governance---when collaboratively designed and contextually adapted---may address long-standing sustainability challenges in OSS projects. This case study demonstrates that reform may deliver operational benefits typically expected in more formally supported industrial initiatives.

For practitioners, this study reinforces the importance of supporting not only technical contributions but also the processes that govern them. For researchers, it contributes to a growing body of empirical work at the intersection of software engineering, organizational studies, and OSS sustainability.

\bibliographystyle{IEEEtran}
\bibliography{software}

\end{document}